# One-step epitaxy of high-mobility La-doped BaSnO$_3$ films by high-pressure magnetron sputtering

Ruyi Zhang,[1,2,†] Xinyan Li,[3,†] Jiachang Bi,[1,2] Shunda Zhang,[1,2] Shaoqin Peng,[1,2] Yang Song,[1,2] Qinghua Zhang,[3,a)] Lin Gu,[3] Junxi Duan,[4,b)] and Yanwei Cao[1,2,c)]

**AFFILIATIONS**

[1]Ningbo Institute of Materials Technology and Engineering, Chinese Academy of Sciences, Ningbo 315201, China

[2]Center of Materials Science and Optoelectronics Engineering, University of Chinese Academy of Sciences, Beijing 100049, China

[3]Beijing National Laboratory for Condensed Matter Physics, Institute of Physics, Chinese Academy of Sciences, Beijing 100190, China

[4]School of Physics, Beijing Institute of Technology, Beijing 100081, China

[†]R. Zhang and X. Li contributed equally to this work.

[a,b,c)]Author to whom correspondence should be addressed: zqh@iphy.ac.cn; junxi.duan@bit.edu.cn; ywcao@nimte.ac.cn

**ABSTRACT**

As a unique perovskite transparent oxide semiconductor, high-mobility La-doped BaSnO$_3$ films have been successfully synthesized by molecular beam epitaxy and pulsed laser deposition. However, it remains a big challenge for magnetron sputtering, a widely applied technique suitable for large-scale fabrication, to grow high-mobility La-doped BaSnO$_3$ films. Here, we developed a method to synthesize high-mobility epitaxial La-doped BaSnO$_3$ films (mobility up to 121 cm$^2$V$^{-1}$s$^{-1}$ at the carrier density ~ 4.0 ×10$^{20}$ cm$^{-3}$ at room temperature) directly on SrTiO$_3$ single crystal substrates using high-pressure magnetron sputtering. The structural and electrical properties of the La-doped BaSnO$_3$ films were characterized by combined high-resolution X-ray diffraction, X-ray photoemission spectroscopy, and temperature-dependent electrical transport measurements. The room temperature electron mobility of La-doped BaSnO$_3$ films in this work is 2 to 4 times higher than the reported values of the films grown by magnetron sputtering. Moreover, in the high carrier density range ($n$ > 3 ×10$^{20}$ cm$^{-3}$), the electron mobility value of 121 cm$^2$V$^{-1}$s$^{-1}$ in our work is among the highest values for all reported doped BaSnO$_3$ films. It is revealed that high argon pressure during sputtering plays a vital role in stabilizing the fully relaxed films and inducing oxygen vacancies, which benefit the high mobility at room temperature. Our work provides an easy and economical way to massively synthesize high-mobility transparent conducting films for transparent electronics.

**INTRODUCTION**

As materials at the heart of transparent electronics, transparent oxide semiconductors (TOS) with a wide band gap and high carrier mobility at room temperature have attracted tremendous interest due to their successful applications in logic devices, solar cells, display panels, and to mention a few.[1-6] In particular, the demonstration of high electron mobility (320 $cm^2V^{-1}s^{-1}$ at room temperature), wide band gap (~ 3.5 eV), and superior high-temperature thermal stability in La-doped $BaSnO_3$ (LBSO) single crystals further boosts the study of perovskite TOS.[7-10] Perovskite oxides with a simple $ABO_3$ chemical formula possess many remarkable properties such as superconductivity,[11,12] piezoelectricity/ferroelectricity,[13,14] and colossal magnetoresistance.[15-17] However, the low carrier mobility at room temperature in perovskite oxides has long been an obstacle for the development of perovskite oxide electronics,[10,18] e.g., the low electron mobility (< 10 $cm^2V^{-1}s^{-1}$) at room temperature in $SrTiO_3$-based compounds and heterostructures.[19-21] The discovery of high-mobility LBSO is a breakthrough and endows a possibility of integrating high-mobility TOS with other perovskite functional oxides for next-generation electronic devices.[22-24]

Despite the very high mobility demonstrated in LBSO single crystals, the mobility of LBSO films is comparatively low. The room temperature electron mobility is only 0.69 $cm^2 V^{-1} s^{-1}$ in early LBSO films grown by pulsed laser deposition (PLD),[25] due to suppression by various structural defects (*i.e.* grain boundaries, misfit dislocations, point defects) in films.[26-28] To decrease the defect density and increase the mobility of LBSO films for applicable devices, several strategies were then developed.[22-43] The threading dislocations (TDs) are almost inevitable in LBSO films due to the lack of substrates with small lattice and symmetry mismatch for the epitaxy of LBSO (*a* ~ 4.12 Å) films.[28] Therefore, on the one hand, less lattice-mismatched substrates such as $DyScO_3$ ($a_{pc}$ = 3.943 Å with lattice mismatch ε = - 4.2 %) instead of $SrTiO_3$ (STO, *a* = 3.905 Å with ε = - 5.2 %) were applied to further increase the mobility of LBSO films.[22,24,31] Up to now, the highest room-temperature mobility of 183 $cm^2V^{-1}s^{-1}$ was reported in epitaxial LBSO films grown on $DyScO_3$(001) substrate by molecular-beam epitaxy (MBE).[24] On the other hand, inserting insulating buffer layers between LBSO films and substrates provides another route to effectively reduce TD density.[33,41,42,44] The strategy of post-annealing was also adopted to increase the room temperature mobility after deposition.[34,37,39] In particular, oxygen vacancies ($V_O$) induced by high-temperature annealing under wet $H_2$ atmosphere or vacuum can accelerate the movement of dislocations and promote lateral grain growth in LBSO films, leading to an increment of room-temperature mobility up to 122 $cm^2V^{-1}s^{-1}$.[37,39] Thus far, almost all reported high-mobility LBSO films were synthesized by MBE and PLD.[22-42] Other deposition techniques (*i.e.,* sputtering, chemical solution deposition) can also grow epitaxial LBSO films, but none of them can reach the crystalline quality and electron mobility as good as those grown by MBE and PLD.[45-50] Given the limited choices of deposition method and strong dependence on the extra steps (*i.e.,* buffer layer, post-annealing) for high-mobility LBSO films, developing an easy and economical technical alternative suitable for large-scale fabrication is very important and appealing for their wide applications in transparent electronics.

As a widely applied deposition technique in both laboratory and industry, magnetron sputtering is remarkable in growing large-size, stoichiometric, and highly homogeneous films. However, at present, the BaSnO$_{3-\delta}$ and LBSO thin films deposited by magnetron sputtering show low room temperature mobility (only 11 - 50 cm$^2$V$^{-1}$s$^{-1}$).[43,46-51] It is noted that the mobility can be heavily suppressed by the low working pressure (< 0.05 mbar) during sputtering due to the high kinetic energy of the plume and inferior crystalline quality of films.[46,49,50] By using high working pressure during the film growth, the kinetic energy of sputtered particles can be significantly decreased, thus reducing their damage to the films and increasing the crystalline qualities.[43,47,48,51,52] However, in these reports, the mobility is still low (< 50 cm$^2$V$^{-1}$s$^{-1}$). Meanwhile, the high argon pressure offers an oxygen-deficient atmosphere during deposition, which can promote V$_O$-assisted lateral grain growth, reduce TD density, and increase the room temperature mobility of LBSO films.[37,39] The combination of high-pressure magnetron sputtering and pure agon atmosphere is expected to significantly improve the crystalline qualities and room temperature mobility of LBSO films.

In this work, we report the one-step epitaxy of high-quality LBSO films by high-pressure magnetron sputtering. The electron mobility can be 121 cm$^2$V$^{-1}$s$^{-1}$ at the carrier density ~ 4.0 ×10$^{20}$ cm$^{-3}$ at room temperature, a record high value in the high carrier density range ($n$ > 3 ×10$^{20}$ cm$^{-3}$) for all reported LBSO films. We characterized the structural and electrical properties of LBSO films by combined high-resolution X-ray diffraction, X-ray photoemission spectroscopy, and temperature-dependent electrical transport measurements. For the first time, we demonstrate that high-pressure argon sputtering can be a promising and alternative technique for the growth of high-mobility LBSO films. More importantly, the one-step epitaxy method demonstrated in this work is free of the buffer layer, post-annealing, and ozone generator, providing an easy and economical way to massively synthesize high-mobility TOS films for transparent electronics.

**EXPERIMENTAL DETAILS**

The two-inch La$_{0.04}$Ba$_{0.96}$SnO$_3$ targets with La-dopants concentration ($n_{nom}$ = 5.74 ×10$^{20}$ cm$^{-3}$) were synthesized by solid-state reaction method with the initial reactants consisted of La$_2$O$_3$ (99.99%), BaCO$_3$ (99.99%), and SnO$_2$ (99.99%). Epitaxial LBSO films were directly deposited on STO (001) substrates (5×5×0.5 mm$^3$) via a home-made radio-frequency (RF) magnetron sputtering system. The base pressure was pumped near 10$^{-6}$ mbar. A high-pressure (0.5 mbar) pure argon atmosphere (99.999%) was maintained during growth with a flow rate of 18 sccm. The RF power was kept at 40 W. The distance between substrate and target was set at 10 cm during sputtering. The deposition rate for LBSO films was around 7.5 Å/min. The LBSO film was uniform on the 1-inch scale substrate with the relative deviation of film thickness distribution less than 5% (see Fig. S1 of the supplementary material). The substrate temperatures ($T_S$) were tested from 750 °C to 900 °C with a step of 50 °C to determine the optimal temperature for the highest mobility in LBSO films. A series of LBSO films with thicknesses ($t$) ranging from 16 - 240 nm were grown at the fixed temperature of 850 °C. To reveal the effect of pure argon atmosphere on the structural and electrical properties, LBSO reference films were grown at 850 °C under the atmosphere of 0.01 mbar $O_2$ and 0.49 mbar Ar mixing gas atmosphere. After growth, all LBSO films immediately followed a natural cooling process to room temperature under the same atmosphere used during deposition without any further annealing. Comparing with the natural cooling process in pure Ar atmosphere, the effects of cooling in a high oxygen environment or low-temperature oxygen post-deposition annealing on the electrical properties of LBSO films can be neglected (see Table S1 of the supplementary material), which is due to the high cooling rate of the heating stage in this work (see Fig. S2 of the supplementary material) and low oxygen diffusion constant (~$10^{-16}$ cm$^2$ s$^{-1}$) of LBSO films.[53] It is important to note that all STO substrates of LBSO/STO films are insulating in this work, excluding the contribution of STO substrates to the conductivity. (see Fig. S3 of the supplementary material).

High-resolution X-ray diffractometer (HRXRD, Bruker D8 Discovery) with monochromatic Cu K$_{\alpha 1}$ radiation wavelength of 1.5406 Å was performed to characterize the structural properties (such as full width at half maximum (FWHM) and lattice parameters) of LBSO films. The surface morphology of LBSO films was scanned by atomic force microscopy (AFM, Bruker Dimension ICON SPM). The transmission electron microscope (TEM) samples were prepared by using Focused Ion Beam (FIB) milling. Cross-sectional lamella was thinned down to 100 nm thick at an accelerating voltage of 30 kV with a decreasing current from the maximum 2.5 nA, followed by fine polish at an accelerating voltage of 2 kV with a small current of 40 pA. The atomic structures of the LBSO films were characterized using an ARM 200CF (JEOL, Tokyo, Japan) transmission electron microscope operated at 200 kV and equipped with double spherical aberration (Cs) correctors. The high-angle annular dark-field (HAADF) images were acquired at an acceptance angle of 68~260 mrad. The chemical compositions (valence states and contents of elements) were detected by X-ray photoemission spectroscopy (XPS, Kratos AXIS Supra) at room temperature with an acceptance angle of 45°. Resistivity, Hall mobility, and carrier density of LBSO films were measured with a van der Pauw geometry by a physical properties measurement system (PPMS, Dynacool QD). Indium electrodes were used at the four corners of LBSO films for ohmic contacts.

**RESULTS AND DISCUSSION**

A series of high-quality LBSO films with thicknesses ranging from 16 to 240 nm were grown on SrTiO$_3$ single crystal substrates (5×5×0.5 mm$^3$). Firstly, to find the optimal growth temperature, several 180 nm-thick LBSO films were grown (at a fixed argon pressure of 0.5 mbar) with substrate temperature $T_S$ varying from 750 °C to 900 °C. As seen in Fig. 1, all LBSO films are highly transparent [Fig. 1(a)] and the most appropriate growth temperature for high electron mobility [121 cm$^2$V$^{-1}$s$^{-1}$, Fig. 1(b)] is in the range 800 - 850 °C. During sputtering, the argon pressure also has a significant effect on the electrical properties of LBSO films (see Fig. S4 of the supplementary material). It is revealed that the pressure $P_{Ar}$ = 0.5 mbar is an optimal condition for carrier mobility ($\mu$) and carrier density ($n$). Then we fixed the growth temperature at 850 °C and argon pressure at 0.5 mbar for the epitaxy of LBSO films with thicknesses ranging from 16 to 240 nm.

Taking 16- and 240-nm thick LBSO films as a representation, we characterized the crystal structure of LBSO films by HRXRD and AFM. As seen in Fig. 2(a), clear thickness fringes around LBSO (002) diffraction peaks can be observed on both films, indicating smooth growth and high crystallinity of thin films. Figure 2(b) shows the crystalline quality of LBSO films and STO substrate evaluated by the FWHM from the rocking curves. As seen, the FWHM decreases from 0.039° for 16 nm to 0.022° for 240 nm, indicating a narrower mosaic spreading in a thicker film. Comparing to the FWHM of 0.016° for STO substrate, the FWHM of 0.022° for 240 nm-thick LBSO film is slightly broader but still on the same level (FWHM ~ 0.02°) of the high-quality LBSO films synthesized by MBE and PLD.[24,42] To characterize the surface morphology of films, we performed AFM [Fig. 2(c)], showing that the root-mean-square roughness (RMS) for 16 nm- and 240 nm-thick LBSO films are 0.39 nm and 0.91 nm, respectively. Reciprocal space mappings (RSMs) measured around (103) diffractions [Fig. 2(d) and Fig. S5 of the supplementary material] not only reveal epitaxial growth of LSBO films on STO substrates but also provide detailed structural information including mosaic spreading, lattice parameters, and in-plane/out-of-plane grain size. All of the RSM peaks for LBSO films with different thicknesses reside at the green reference lines indicating a cubic symmetry. The lattice parameters for all films equal ~ 4.122 Å, slightly larger than the average lattice parameters 4.117 Å of the $La_{0.04}Ba_{0.96}SnO_3$ film.[28] It seems common that the lattice parameters of $BaSnO_3$ films in literature are smaller than what we have observed here.[22,24] However, many research also observed unusual slight expansion of lattice parameters on STO substrates, which were ascribed to minor cation non-stoichiometry or defect formation.[47] Here, the slight expansion of lattice parameters is mainly attributed to $V_O$ doping, which has also been observed in doped and undoped BSO films and single crystals.[47,54,55] Unlike partially strained states observed for LBSO/STO (001) films with thicknesses below 100 nm in the previous study,[28,36] the strain in our films quickly gets fully relaxed even with the minimum thickness of 16 nm. Then, we estimate the lateral grain sizes ($D$) of LBSO films by using the formula $D = 1/\Delta Q_x$, where $\Delta Q_x$ is the integral width in the $Q_x$ direction of RSMs.[36] The integral width equals the FWHM of the projection curve obtained by adding the intensities of RSM data point with the same $Q_x$ coordinates. As seen in Fig. S6 of the supplementary material, $D$ ($\Delta Q_x$) increases (decreases) from 31.1 nm (0.0322 nm$^{-1}$) for 16 nm thick film to 86.2 nm (0.0116 nm$^{-1}$) for 240 nm thick film, further confirming improved crystallinity in thicker films.

To reveal the microstructure of LBSO thin films, we carried out TEM characterizations. Figure 3 shows the TEM images of a 240 nm- thick LBSO film grown in the pure argon atmosphere. As seen in Fig. 3(a), periodic misfit dislocations (MD) with an average distance of 6 nm can be clearly observed at the interface between LBSO film and STO substrate. The MD spacing of 6 nm is smaller than the calculated value of 7.5 nm and reported values (7 ~ 8 nm) for the LBSO/STO films.[3,36,47,56]

The HAADF-TEM image [see Fig. 3 (b)] shows the interface is atomically sharp, and the LBSO film is of high crystalline quality. The TEM bright-field image in Fig. 3 (c) shows the distribution of TDs (yellow bars) near the surface. The average distance between two TDs is ~ 80 nm. The estimated TD density is around $1.6 \times 10^{10}$ cm$^{-2}$, which is smaller than previously reported values of LBSO/STO films.[28,54,56] The LBSO film in the HAADF-TEM image also shows less columnar structure than those in literature.[28,54,56] According to the remarkably low TD density and less columnar structure, a high mobility is expected in this work. It is noted that some TDs stop expanding at the positions away from the interfaces (see blue arrows), indicating the TD density depends on the thickness of LBSO film. As shown in Fig. S7 of the supplementary material, the average distance ($L$) of TDs increases, whereas the TD density (N) decreases with the increase of the position ($l$) from the interface between the film and substrate. This behavior agrees well with the feature shown by the thickness ($t$) dependent grain size ($D$), which is extracted from the RSM integral width analyses. Therefore, thicker LBSO films have lower TD density and larger grain size, which can benefit high mobility at room temperature.

To investigate the electrical properties of LBSO films, we extracted thickness- and temperature-dependent carrier mobility and density from electrical measurements. As seen in Fig. 4(a) and (b), the electron mobilities (densities) are 73 cm$^2$V$^{-1}$s$^{-1}$ ($3.0 \times 10^{20}$ cm$^{-3}$), 77 cm$^2$V$^{-1}$s$^{-1}$ ($3.3 \times 10^{20}$ cm$^{-3}$), 97 cm$^2$V$^{-1}$s$^{-1}$ ($3.8 \times 10^{20}$ cm$^{-3}$), 121 cm$^2$V$^{-1}$s$^{-1}$ ($4.0 \times 10^{20}$ cm$^{-3}$), and 120 cm$^2$V$^{-1}$s$^{-1}$ ($4.1 \times 10^{20}$ cm$^{-3}$) for 16 nm-, 28 nm-, 90 nm-, 180 nm-, and 240 nm-thick LBSO films, respectively. The saturated electron density (~ $4.0 \times 10^{20}$ cm$^{-3}$) from Hall measurements corresponds to an activation ratio of ~70 % of the nominal value of 4% La concentration ($5.74 \times 10^{20}$ cm$^{-3}$). The critical thickness ($t_C$) for saturation is near 180 nm. Similar saturation behavior for the room temperature carrier mobility and density with increasing thickness has been observed elsewhere.[36,57,58] The critical thickness here is only half of the reported $t_C$ (~ 400 nm) in literature.[36] The structural defects including TDs not only act as scattering sources but also as trap centers for charges, which can decrease the carrier mobility and density simultaneously.[28] The thickness-dependent electrical properties observed here are highly connected with the improved crystallinity in the thicker film as evidenced from Fig. 2&3 and Fig. S6&S7 of the supplementary material. Next, we study the temperature-dependent electrical properties of 180 nm-thick LBSO film. As shown in Fig. 4 (c), the resistivity ($\rho$) decreases with cooling down, showing a metallic behavior with high conductivity of 15100 S·cm$^{-1}$ at 2 K and 7700 S·cm$^{-1}$ at 300 K. On the other hand, the electron mobility increases from 121 cm$^2$V$^{-1}$s$^{-1}$ at 300 K to 237 cm$^2$V$^{-1}$s$^{-1}$ at 2 K [Fig. 4(d)], whereas the electron density (~ $4.0 \times 10^{20}$ cm$^{-3}$) is almost independent of temperature [Fig. 4(e)], a typical feature of a degenerate semiconductor.[22,24] The overall tendency shown in Fig. 4(d) and 4(e) is in good consistency with the result from high-mobility LBSO films in literature.[22,24]

To understand the effect of deposition atmosphere on the mobility of LBSO films, LBSO

reference films were synthesized in an Ar/O$_2$ mixing atmosphere (0.49 mbar Ar and 0.01 mbar O$_2$, labeled as LBSO[Ar+O$_2$]). Comparing to the LBSO films grown in pure Ar atmosphere (labeled as LBSO[Ar]), the growth condition of LBSO[Ar+O$_2$] films was kept the same except for the oxygen partial pressure. Hall effect measurements at room temperature show that electron mobility and density of 180 nm- thick LBSO[Ar+O$_2$] films are only 74.0 cm$^2$V$^{-1}$s$^{-1}$ and 0.8 ×10$^{20}$ cm$^{-3}$, respectively, which are much lower than the values of LBSO[Ar] films. The RSM of LBSO[Ar+O$_2$] films in Fig. S8(a) of the supplementary material reveals the coexistence of partially strained phase and fully relaxed phase, a contrast to the pure fully relaxed phase in LBSO[Ar] films. Further analyses (see Fig. S8(b) of the supplementary material) show that the integral width is much larger ($\Delta Q_x$ = 0.0486 nm$^{-1}$ with grain size $D$ of 20.6 nm) in LBSO[Ar+O$_2$] films than that ($\Delta Q_x$= 0.0131 nm$^{-1}$ with grain size $D$ of 76.3 nm) in LBSO[Ar] films, indicating higher TD density in LBSO[Ar+O$_2$] films. The TEM bright-field image of the LBSO[Ar+O$_2$] films (see Fig. S9 of the supplementary material) shows a higher TD density (~ 1.5×10$^{11}$ cm$^{-2}$) with an average distance of ~ 26 nm between two TDs, consistent with the analysis of RSM. It is revealed that the TD density of partially strained phases is higher, whereas fully relaxed phases benefit lower TD density. It is noted that a previous study on LBSO films has indicated that the lattice strain can hinder grain growth and TD annihilation.[36,39] Therefore, the pure argon atmosphere during sputtering plays a vital role in stabilizing the fully relaxed phases of LBSO films which benefit lower TD density and higher Hall mobility at room temperature.

To evaluate V$_O$ density in LBSO films prepared under different atmospheres, we carried out the O1s XPS measurement on LBSO[Ar] and LBSO[Ar+O$_2$] films. The XPS La 3d, Ba 3d, and Sn 3d spectra of the LBSO [Ar] films are shown in Fig. S10 of the supplementary material. As seen in Fig. 5, both O1s spectra of LBSO[Ar] and LBSO[Ar+O$_2$] films can be deconvoluted into lattice oxygen (~529.6 eV, denoted as O$_L$), chemically adsorbed oxygen (~532.1 eV), and V$_O$ (530.9 eV).[39,59] The V$_O$/O$_L$ area ratio is 0.8 in LBSO[Ar] films, whereas the V$_O$/O$_L$ area ratio is only 0.12 in LBSO[Ar+O$_2$] films, indicating much higher V$_O$ density in LBSO[Ar] films. It can be concluded that a high-pressure Ar atmosphere favors V$_O$ generation, stabilizes the fully relaxed state, and reduces TD density, leading to high room temperature mobility. Previously, the V$_O$-assisted recovery process has been demonstrated in LBSO films by post-annealing under the wet H$_2$ atmosphere or vacuum.[37,39] V$_O$ can accelerate TDs movement across the slip plane and annihilate part of TDs.[37] It has also been pointed out that point defects like V$_O$ can help to release deformed strain,[60,61] which supports our observation. The smaller MD spacing of 6 nm observed in this work also indicates that V$_O$ can participate in the early stage of strain relaxation near the film-substrate interface.

Next, we summarize the room temperature electrical properties of LBSO films here in this work and those in literature. As seen in Fig. 6(a), the room temperature mobility 121 cm$^2$V$^{-1}$s$^{-1}$ here in this work is 2 to 4 times higher than that of LBSO films grown by sputtering in literature.[43,46,49,50]

Moreover, in the high carrier density range ($n > 3 \times 10^{20}$ cm$^{-3}$) the room temperature mobility 121 cm$^2$V$^{-1}$s$^{-1}$ in our work is among the highest values in the record.[22-50,57-61] The best room temperature conductivity of 7700 S·cm$^{-1}$ in this work is also among the highest values in the carrier density range ($\leq 8.0 \times 10^{20}$ cm$^{-3}$) in record [Fig. 6(b)].[22-50,57-61] A detailed comparison of room temperature carrier mobility, carrier density, conductivity, and growth methods of LBSO films in literature can also be referred to Table. S2 of the supplementary material. It is noted that the method of high-pressure argon sputtering to promote Hall mobility is not only effective on STO substrates but also applicable for other widely used perovskite substrates, e.g., LaAlO$_3$ (LAO), TbScO$_3$ (TSO), and KTaO$_3$ (KTO) substrates (see Fig. S11 of the supplementary material). All LBSO films deposited on these substates show high room temperature mobility over 100 cm$^2$V$^{-1}$s$^{-1}$. The LBSO films on KTO substrates even show slightly higher room temperature electrical parameters (122 cm$^2$V$^{-1}$s$^{-1}$ at 4.3 $\times 10^{20}$ cm$^{-3}$) than those on STO substrates. High-temperature annealing in air or vacuum can move oxygen in or out of LBSO films, leading to tunable structural and electrical properties (see Fig. S12 of the supplementary material). The significant advance of room temperature mobility shown in this work demonstrates that high-pressure sputtering can also be a very appealing technique for high-quality TOS film synthesis.

## SUMMARY

In summary, we synthesized a series of high-mobility LBSO films by high-pressure magnetron sputtering. The electron mobility can reach 121 cm$^2$V$^{-1}$s$^{-1}$ at the carrier density ~ 4.0 $\times 10^{20}$ cm$^{-3}$ at room temperature, a record high value in the high carrier density range ($n > 3 \times 10^{20}$ cm$^{-3}$) for all reported LBSO films. The structural and electrical properties of LBSO films were characterized by combined high-resolution X-ray diffraction, X-ray photoemission spectroscopy, and temperature-dependent electrical transport. It is revealed that high argon pressure during sputtering plays a vital role in stabilizing the films and induing oxygen vacancies, which benefits high-mobility. For the first time, we demonstrate that high-pressure argon sputtering can be a promising and alternative technique for the growth of high-mobility LBSO films. More importantly, the method of one-step epitaxy shown in this work is free of the buffer layer, post-annealing, and ozone generator. Our work provides an easy and economical way to massively synthesize high-mobility transparent conducting films for transparent electronics.

## SUPPLEMENTARY MATERIAL

See the supplementary material for Photo image showing sputtering deposition; Cooling rate for the heating stage; Argon pressure dependent electrical transport properties and RSMs with different thickness for LBSO films; Integral width analyses; XPS spectra for LBSO thin film; TEM image for LBSO film grown in mixing Ar/O$_2$ atmosphere; Comparison of RSMs of LBSO films grown under different atmosphere; Substrate dependent and annealing tunable structural and electrical transport properties for LBSO films.

## AUTHORS' CONTRIBUTIONS

R. Zhang and Y. Cao conceived the project. R. Zhang prepared the samples. R. Zhang, J. Bi, S. Peng, and Y. Song performed the HRXRD, AFM, and XPS characterizations. X. Li, Q. Zhang, and L. Gu performed the TEM characterizations and analyses. J. Duan and R. Zhang measured the electrical transport properties. All authors discussed the data and contributed to the manuscript.

## ACKNOWLEDGMENTS

The authors deeply acknowledge insightful discussions with Fang Yang and Jiandong Guo. This work was supported by the National Natural Science Foundation of China (Grant Nos. 11874058, 61804008, and U2032126), the Pioneer Hundred Talents Program of Chinese Academy of Sciences, the Natural Science Foundation of Zhejiang Province, the Beijing National Laboratory for Condensed Matter Physics, the Ningbo 3315 Innovation Team, and the Ningbo Science and Technology Bureau (Grant No. 2018B10060). This work was partially supported by Youth Program of National Natural Science Foundation of China (Grant No. 12004399), China Postdoctoral Science Foundation (Grant No. 2018M642500), Postdoctoral Science Foundation of Zhejiang Province (Grant No. zj20180048), and Natural Science Foundation of Ningbo City (Grant No. 202003N4364).

## DATA AVAILABILITY

The data that support the findings of this study are available within the article, its supplementary material, or from the corresponding author upon reasonable request.

## REFERENCES

[1]K. Nomura, H. Ohta, A. Takagi, T. Kamiya, M. Hirano, and H. Hosono, Nature **432**, 488 (2004).
[2]K. Ellmer, Nat. Photonics **6**, 808 (2012).
[3]S. C. Dixon, D. O. Scanlon, C. J. Carmalt, and I. P. Parkin, J. Mater. Chem. C **4**, 6946 (2016).
[4]M. Morales‑Masis, S. D. Wolf, R. Woods‑Robinson, J. W. Ager, and C. Ballif, Adv. Electron. Mater. **3**, 1600529 (2017).
[5]D. Li, W.-Y. Lai, Y.-Z. Zhang, and W. Huang, Adv. Mater. **30**, 1704738 (2018).
[6]A. Prakash and B. Jalan, Adv. Mater. Interfaces **6**, 1900479 (2019).
[7]H. J. Kim, U. Kim, H. M. Kim, T. H. Kim, H. S. Mun, B.-G. Jeon, K. T. Hong, W.-J. Lee, C. Ju, K. H. Kim, and K. Char, Appl. Phys. Express **5**, 061102 (2012).
[8]S. Ismail-Beigi, F. J. Walker, S.-W. Cheong, K. M. Rabe, and C. H. Ahn, APL Mater. **3**, 062510 (2015).
[9]W. J. Lee, H. J. Kim, J. Kang, H. J. Dong, H. K. Tai, J. H. Lee, and K. H. Kim, Annu. Rev. Mater. Res. **47**, 391 (2017).
[10]H. He, Z. Yang, Y. Xu, A. T. Smith, G. Yang, and L. Sun, Nano Convergence **7**, 32 (2020).
[11]N. Reyren, S. Thiel, A. D. Caviglia, L. F. Kourkoutis, G. Hammerl, C. Richter, C. W. Schneider, T.

Kopp, A. S. Ruetschi, D. Jaccard, M. Gabay, D. A. Muller, J. M. Triscone, and J. Mannhart, Science **317**, 1196 (2007).

[12] K. Ueno, S. Nakamura, H. Shimotani, A. Ohtomo, N. Kimura, T. Nojima, H. Aoki, Y. Iwasa, and M. Kawasaki, Nat. Mater. **7**, 855 (2008).

[13] F. Li, D. Lin, Z. Chen, Z. Cheng, J. Wang, C. Li, Z. Xu, Q. Huang, X. Liao, L.-Q. Chen, T. R. Shrout, and S. Zhang, Nat. Mater. **17**, 349 (2018).

[14] S. Das, Y. L. Tang, Z. Hong, M. A. P. Goncalves, M. R. McCarter, C. Klewe, K. X. Nguyen, F. Gomez-Ortiz, P. Shafer, E. Arenholz, V. A. Stoica, S. L. Hsu, B. Wang, C. Ophus, J. F. Liu, C. T. Nelson, S. Saremi, B. Prasad, A. B. Mei, D. G. Schlom, J. Iniguez, P. Garcia-Fernandez, D. A. Muller, L. Q. Chen, J. Junquera, L. W. Martin, and R. Ramesh, Nature **568**, 368 (2019).

[15] Y. Moritomo, A. Asamitsu, H. Kuwahara, and Y. Tokura, Nature **380**, 141 (1996).

[16] G. M. Zhao, K. Conder, H. Keller, and K. A. Muller, Nature **381**, 676 (1996).

[17] K. L. Kobayashi, T. Kimura, H. Sawada, K. Terakura, and Y. Tokura, Nature **395**, 677 (1998).

[18] F. Trier, D. V. Christensen, and N. Pryds, J. Phys. D: Appl. Phys. **51**, 293002 (2018).

[19] A. Ohtomo and H. Y. Hwang, Nature **427**, 423 (2004).

[20] S. Thiel, G. Hammerl, A. Schmehl, C. W. Schneider, and J. Mannhart, Science **313**, 1942 (2006).

[21] J. Son, P. Moetakef, B. Jalan, O. Bierwagen, N. J. Wright, R. Engel-Herbert, and S. Stemmer, Nat. Mater. **9**, 482 (2010).

[22] S. Raghavan, T. Schumann, H. Kim, J. Y. Zhang, T. A. Cain, and S. Stemmer, APL Mater. **4**, 016106 (2016).

[23] S. Heo, D. Yoon, S. Yu, J. Son, and H. M. Jang, J. Mater. Chem. C **5**, 11763 (2017).

[24] H. Paik, Z. Chen, E. Lochocki, A. H. Seidner, A. Verma, N. Tanen, J. Park, M. Uchida, S. Shang, B.-C. Zhou, M. Brutzam, R. Uecker, Z.-K. Liu, D. Jena, K. M. Shen, D. A. Muller, and D. G. Schlom, APL Mater. **5**, 116107 (2017).

[25] H. F. Wang, Q. Z. Liu, F. Chen, G. Y. Gao, W. Wu, and X. H. Chen, J. Appl. Phys. **101**, 106105 (2007).

[26] H. J. Kim, U. Kim, T. H. Kim, J. Kim, H. M. Kim, B.-G. Jeon, W.-J. Lee, H. S. Mun, K. T. Hong, J. Yu, K. Char, and K. H. Kim, Phys. Rev. B **86**, 165205 (2012).

[27] Q. Liu, J. Liu, B. Li, H. Li, G. Zhu, K. Dai, Z. Liu, P. Zhang, and J. Dai, Appl. Phys. Lett. **101**, 241901 (2012).

[28] H. Mun, U. Kim, H. M. Kim, C. Park, T. H. Kim, H. J. Kim, K. H. Kim, and K. Char, Appl. Phys. Lett. **102**, 252105 (2013).

[29] C. Park, U. Kim, C. J. Ju, J. S. Park, Y. M. Kim, and K. Char, Appl. Phys. Lett. **105**, 203503 (2014).

[30] U. Kim, C. Park, T. Ha, Y. M. Kim, N. Kim, C. Ju, J. Park, J. Yu, J. H. Kim, and K. Char, APL Mater. **3**, 036101 (2015).

[31] W.-J. Lee, H. J. Kim, E. Sohn, T. H. Kim, J.-Y. Park, W. Park, H. Jeong, T. Lee, J. H. Kim, K.-Y. Choi, and K. H. Kim, Appl. Phys. Lett. **108**, 082105 (2016).

[32] C. A. Niedermeier, S. Rhode, S. Fearn, K. Ide, M. A. Moram, H. Hiramatsu, H. Hosono, and T. Kamiya, Appl. Phys. Lett. **108**, 172101 (2016).

[33] J. Shin, Y. M. Kim, Y. Kim, C. Park, and K. Char, Appl. Phys. Lett. **109**, 262102 (2016).

[34] S. Yu, D. Yoon, and J. Son, Appl. Phys. Lett. **108**, 262101 (2016).

[35] K. Fujiwara, K. Nishihara, J. Shiogai, and A. Tsukazaki, Appl. Phys. Lett. **110**, 203503 (2017).

[36] A. V. Sanchela, M. Wei, H. Zensyo, B. Feng, J. Lee, G. Kim, H. Jeen, Y. Ikuhara, and H. Ohta, Appl. Phys. Lett. **112**, 232102 (2018).

[37] D. Yoon, S. Yu, and J. Son, NPG Asia Mater. **10**, 363 (2018).

[38]U. S. Alaan, F. J. Wong, J. J. Ditto, A. W. Robertson, E. Lindgren, A. Prakash, G. Haugstad, P. Shafer, A. T. N'Diaye, D. Johnson, E. Arenholz, B. Jalan, N. D. Browning, and Y. Suzuki, Phys. Rev. Mater. **3**, 124402 (2019).

[39]H. J. Cho, T. Onozato, M. Wei, A. Sanchela, and H. Ohta, APL Mater. **7**, 022507 (2019).

[40]A. V. Sanchela, M. Wei, J. Lee, G. Kim, H. Jeen, B. Feng, Y. Ikuhara, H. J. Cho, and H. Ohta, J. Mater. Chem. C **7**, 5797 (2019).

[41]A. P. N. Tchiomo, W. Braun, B. P. Doyle, W. Sigle, P. van Aken, J. Mannhart, and P. Ngabonziza, APL Mater. **7**, 041119 (2019).

[42]Z. Wang, H. Paik, Z. Chen, D. A. Muller, and D. G. Schlom, APL Mater. **7**, 022520 (2019).

[43]H. Wang, A. Prakash, K. Reich, K. Ganguly, B. Jalan, and C. Leighton, APL Mater. **8**, 071113 (2020).

[44]A. Prakash, P. Xu, A. Faghaninia, S. Shukla, J. W. Ager, 3rd, C. S. Lo, and B. Jalan, Nat. Commun. **8**, 15167 (2017).

[45]R. H. Wei, X. W. Tang, Z. Z. Hui, X. Luo, J. M. Dai, J. Yang, W. H. Song, L. Chen, X. G. Zhu, X. B. Zhu, and Y. P. Sun, Appl. Phys. Lett. **106**, 101906 (2015).

[46]X. Fei, B.-C. Luo, K.-X. Jin, and C.-L. Chen, Acta Phys. Sin. **64**, 207303 (2015).

[47]K. Ganguly, P. Ambwani, P. Xu, J. S. Jeong, K. A. Mkhoyan, C. Leighton, and B. Jalan, APL Mater. **3**, 062509 (2015).

[48]K. Ganguly, A. Prakash, B. Jalan, and C. Leighton, APL Mater. **5**, 056102 (2017).

[49]B. Luo and J. Hu, ACS Appl. Electron. Mater. **1**, 51 (2019).

[50]A. Tiwari and M.-S. Wong, Thin Solid Films **703**, 137986 (2020).

[51]H. Wang, J. Walter, K. Ganguly, B. Yu, G. Yu, Z. Zhang, H. Zhou, H. Fu, M. Greven, and C. Leighton, Phys. Rev. Mater. **3**, 075001 (2019).

[52]F. O. L. Johansson, P. Ahlberg, U. Jansson, S.-L. Zhang, A. Lindblad, and T. Nyberg, Appl. Phys. Lett. **110**, 091601 (2017).

[53]W.-J. Lee, H. J. Kim, E. Sohn, H. M. Kim, T. H. Kim, K. Char, J. H. Kim, and K. H. Kim, Phys. Status Solidi A **212**, 1487 (2015).

[54]J. Cui, Y. Zhang, J. Wang, Z. Zhao, H. Huang, W. Zou, M. Yang, R. Peng, W. Yan, Q. Huang, Z. Fu, and Y. Lu, Phys. Rev. B **100**, 165312 (2019).

[55]E. McCalla, D. Phelan, M. J. Krogstad, B. Dabrowski, and C. Leighton, Phys. Rev. Mater. **2**, 084601 (2018).

[56]W. M. Postiglione, K. Ganguly, H. Yun, J. S. Jeong, A. Jacobson, L. Borgeson, B. Jalan, K. A. Mkhoyan, and C. Leighton, Phys. Rev. Mater. **5**, 044604 (2021).

[57]Q. Liu, F. Jin, J. Dai, B. Li, L. Geng, and J. Liu, Superlattices Microstruct. **96**, 205 (2016).

[58]A. Kumar, S. Maurya, S. Chawla, S. Patwardhan, and B. Kavaipatti, Appl. Phys. Lett. **114**, 212103 (2019).

[59]H. J. Cho, B. Feng, T. Onozato, M. Wei, A. V. Sanchela, Y. Ikuhara, and H. Ohta, Phys. Rev. Mater. **3**, 094601 (2019).

[60]G. Lu and E. Kaxiras, Phys. Rev. Lett. **89**, 105501 (2002).

[61]K. Terai, M. Lippmaa, P. Ahmet, T. Chikyow, T. Fujii, H. Koinuma, and M. Kawasaki, Appl. Phys. Lett. **80**, 4437 (2002).

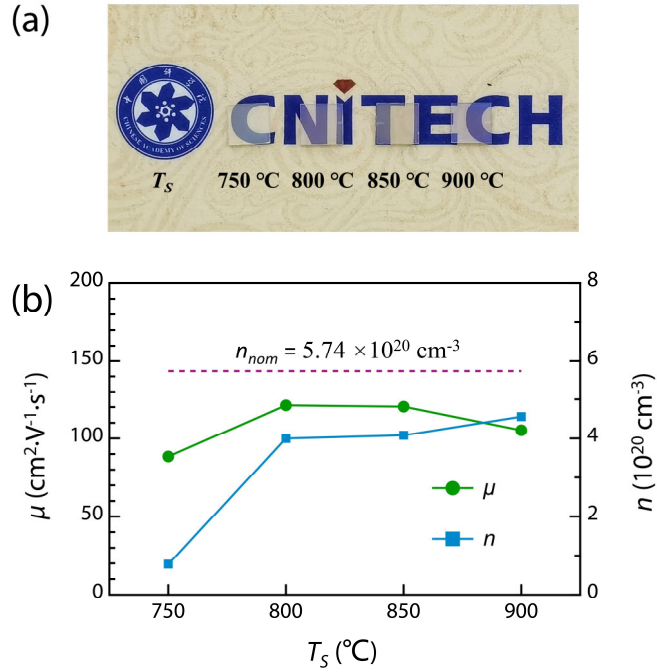

**FIG. 1** (a) Photograph of 180 nm- thick LBSO films grown with various substrate temperature ($T_S$) from 750 °C to 900 °C. (b) $T_S$ - dependent room temperature Hall mobility ($\mu$) and carrier density ($n$) of LBSO films at room temperature.

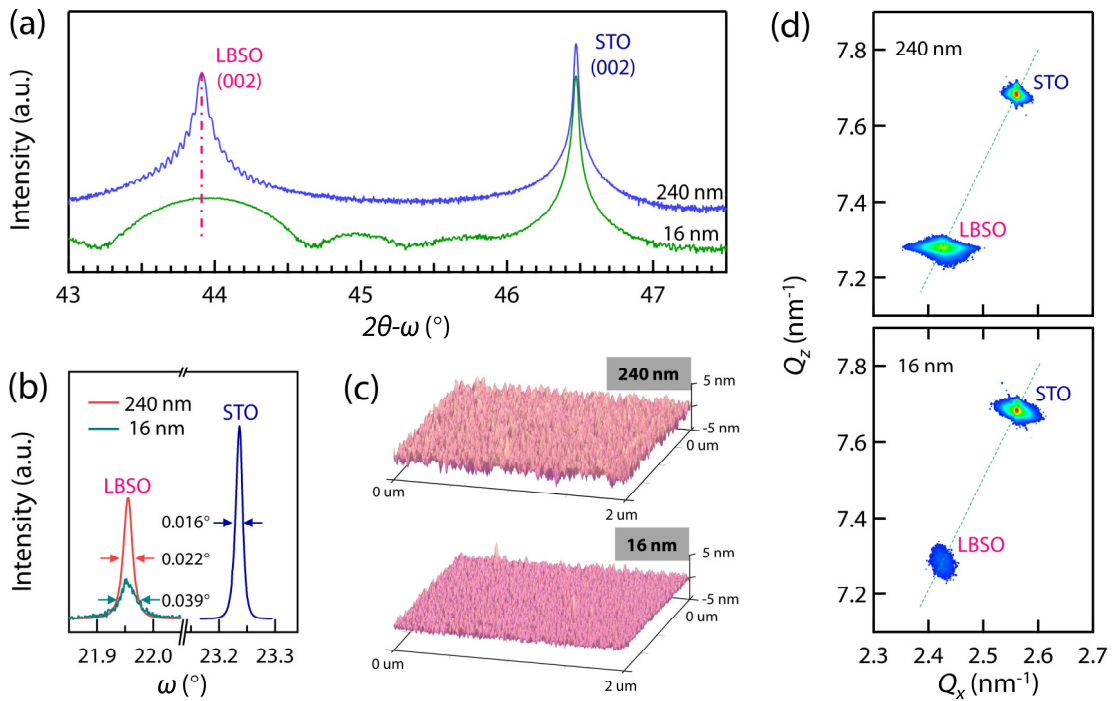

**FIG. 2** Crystal structure of LBSO films (16 nm- and 240 nm-thick) grown on STO (001) substrates. (a) The $2\theta$-$\omega$ scans around (002) diffractions. (b) Rocking curves around (002) diffractions of films and substrates. The arrows indicate the FWHM of films and substrate. (c) AFM surface morphology. (d) RSMs around (103) diffractions. The green reference lines ($Q_z = 3Q_x$) indicate a cubic symmetry.

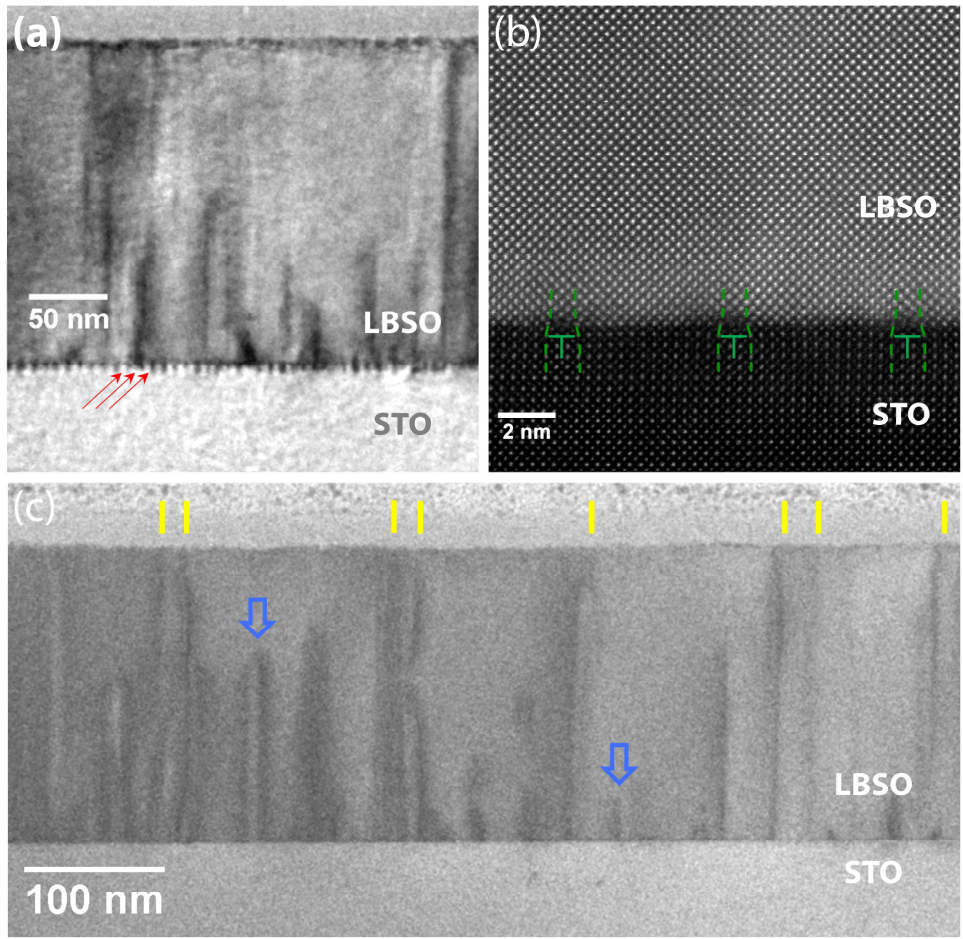

**FIG. 3** TEM images of the 240 nm- thick LBSO film on STO (001) substrate. The low magnification TEM image (a) and HAADF-TEM image (b) showing the periodic misfit dislocations (marked by the red arrows and green edge dislocation symbols) at the interface between the LBSO film and the STO substrate. (c) The TEM bright-field image with threading dislocations (TDs). The TDs near the film surface are labeled with yellow bars. Blue arrows indicate some TDs stop expanding at the positions away from the interfaces.

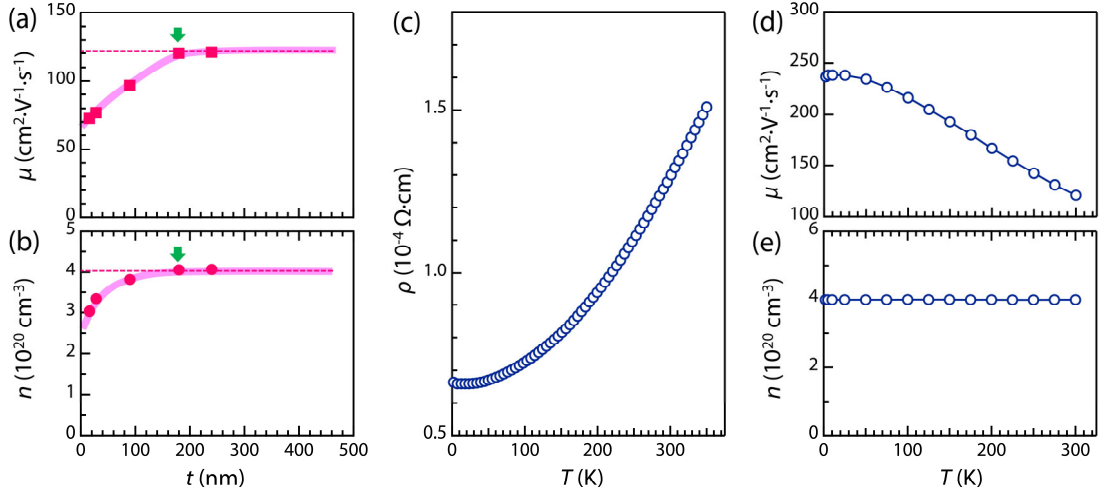

**FIG. 4** (a) and (b) Thickness-dependent Hall mobility ($\mu$) and carrier density ($n$) of LBSO films at room temperature. The green arrows indicate the critical thickness $t_C$, where both $\mu$ and $n$ start to saturate. (c), (d), and (e) Temperature-dependent resistivity ($\rho$), Hall mobility ($\mu$), and carrier density ($n$) of 180 nm thick LBSO films, respectively.

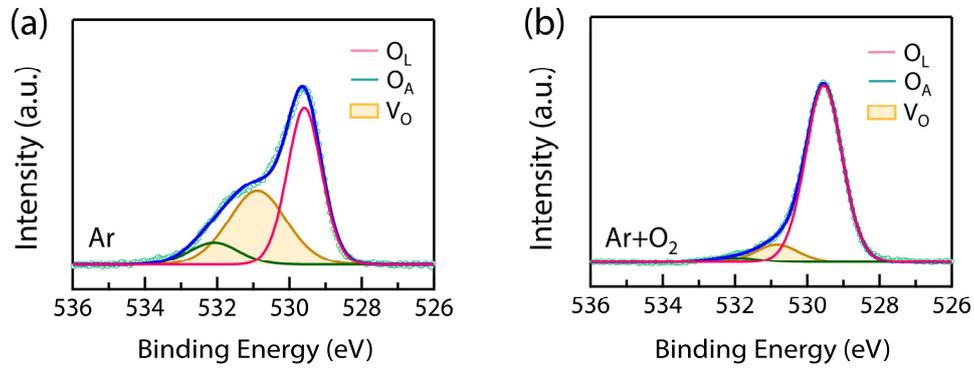

**FIG. 5** Oxygen XPS spectra of LBSO films grown under (a) pure Ar atmosphere (0.5 mBar) and (b) Ar + $O_2$ mixing atmosphere (0.49 mbar Ar + 0.01 mBar $O_2$). Here, $O_L$, $O_A$, and $V_O$ label lattice oxygen, adsorbed oxygen, and oxygen vacancy, respectively.

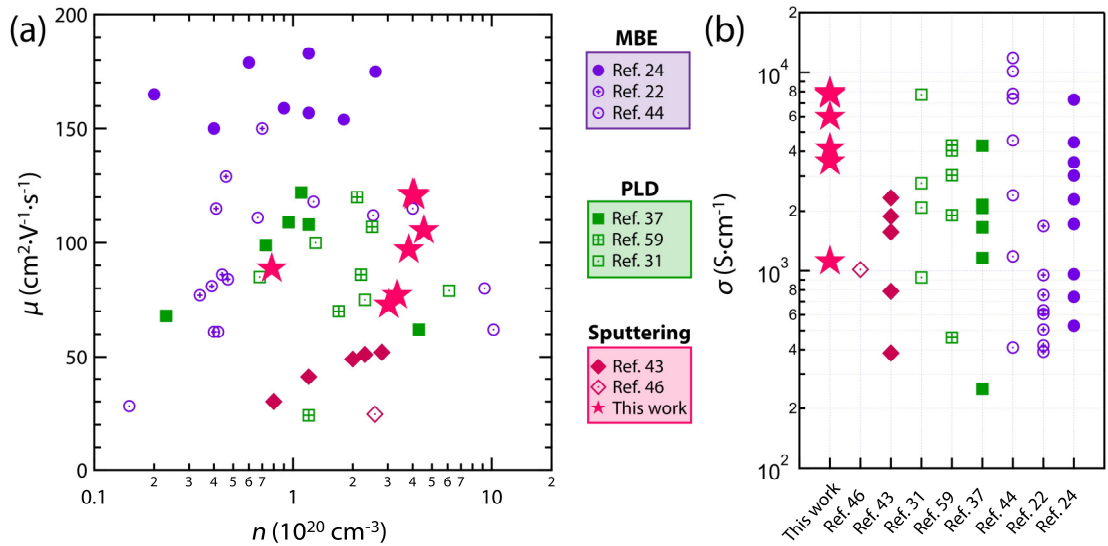

**FIG. 6** Summary of (a) room temperature Hall mobility ($\mu$) and (b) electrical conductivity ($\sigma$) of LBSO films in literature and this work.